\documentclass[twocolumn,nofootinbib,amsmath,amssymb,aps,prd,balancelastpage]{revtex4-1}

\usepackage{graphicx}
\usepackage[caption=false]{subfig}

\usepackage{amsmath,amssymb}
\usepackage{amsfonts,amssymb,mathrsfs}

\usepackage[bookmarks=false,pdfstartview=FitH]{hyperref}
\usepackage[all]{hypcap}

\def\be{\begin{equation}}
\def\ee{\end{equation}}
\def\nn{\nonumber}
\def\f{\frac}
\def\tf{\tfrac}
\def\pl{{\rm Pl}}
\def\lp{\ell_\pl}
\def\b{\bar}

\def\t{\tilde}
\def\v{\vec}
\def\wh{\widehat}

\def\de{\delta}

\def\om{\omega}

\def\mR{\mathcal{R}}
\def\mH{\mathcal{H}}
\def\mO{\mathcal{O}}

\def\mC{\mathcal{C}}

\def\bra{\langle}
\def\ket{\rangle}

\usepackage{color}

\begin{document}

\pagestyle{plain}

\title{Separate universes in loop quantum cosmology: framework and applications}

\author{Edward Wilson-Ewing} \email{wilson-ewing@aei.mpg.de}
\affiliation{Max Planck Institute for Gravitational Physics (Albert Einstein Institute),\\
Am M\"uhlenberg 1, 14476 Golm, Germany, EU}

\begin{abstract}

I present a streamlined review of how the separate universe approach to cosmological perturbation theory can be used to study the dynamics of long-wavelength scalar perturbations in loop quantum cosmology, and then use it to calculate how long-wavelength curvature perturbations evolve across the loop quantum cosmology bounce assuming a constant equation of state.  A similar calculation is possible for tensor modes using results from a complementary approach to cosmological perturbation theory in loop quantum cosmology based on an effective Hamiltonian constraint.  An interesting result is that the tensor-to-scalar ratio can be suppressed or amplified by quantum gravity effects during the bounce, depending on the equation of state of the matter field dominating the dynamics.  In particular, if the equation of state lies between $-1/3$ and $1$, the value of the tensor-to-scalar ratio will be suppressed during the bounce, in some cases significantly.

\end{abstract}

\maketitle

\section{Introduction}
\label{s.intro}

One of the main problems in the search for a theory of quantum gravity is the current lack of any experiments that could offer some guidance.  This is not surprising, since quantum gravity effects are typically only expected to become important either at very high energy scales of the order of the Planck energy $E_p = \sqrt{c^5 \hbar / G} \sim 10^{19}$ GeV, well out of the range of particle colliders and even cosmic rays, or when a space-time curvature scalar (for example the Ricci scalar) becomes sufficiently large, of the order of the Planck curvature $R_p = c^3 / G \hbar \sim 10^{70}$ m${}^{-2}$.  However, there are two physical settings of particular interest where the space-time curvature is expected to become sufficiently large for quantum gravity effects to become important and where, at least in principle, experimental tests of quantum gravity may be possible: these are near the center of black holes and in the very early universe.

Since the high-curvature regime inside astrophysical black holes lies well inside their horizon, it seems that the best hope to obtain any experimental guidance for quantum gravity is via precision observations of the very early universe.  Indeed, the recent results of the Planck collaboration measuring the temperature anisotropies and the polarization of the cosmic microwave background (CMB) \cite{Ade:2015xua} show that this may be possible.  While these results have mostly been used so far to constrain various models of inflation (see, e.g., \cite{Martin:2013tda}) and alternatives to inflation, it is possible that there may be some subleading effects directly due to quantum gravity effects that could also be detected.

In order to check this possibility, it is important to understand the predictions of various quantum gravity theories in their cosmological sector.  One approach to studying quantum gravity effects in cosmology is loop quantum cosmology (LQC), which is based on the minisuperspace quantization of homogeneous cosmological space-times using the techniques of loop quantum gravity.  One of the most interesting results of LQC is that the big-bang and big-crunch singularities, which arise generically in classical general relativity, are resolved due to quantum gravity effects.  Instead, in a contracting semi-classical cosmological space-time, while the dynamics given by classical general relativity can be trusted so long as no curvature invariants are of the order of the Planck scale, once the curvature nears the Planck scale quantum gravity effects become important and a bounce occurs.  Hence, a contracting space-time smoothly bounces to an expanding space-time due to quantum gravity effects in LQC.  The physics and mathematics of LQC are more thoroughly reviewed in \cite{Bojowald:2008zzb, Ashtekar:2011ni, Banerjee:2011qu}.

In order to compare the predictions of LQC with the observed temperature anisotropies of the CMB, it is necessary to be able to determine the dynamics of cosmological perturbations when quantum gravity effects become important.  There currently exist three main proposals on how this can be done.  The first to be developed is an effective theory which aims to capture the leading quantum gravity effects by modifying the classical Hamiltonian and diffeomorphism constraints in a specific manner motivated by LQC while ensuring that no anomalies appear in the (now modified) constraint algebra \cite{Bojowald:2008jv, Cailleteau:2011kr, Cailleteau:2012fy}.  More recently, a hybrid quantization has been proposed, where the homogeneous background is quantized following LQC while the linear perturbations are quantized in the standard Fock manner \cite{FernandezMendez:2012vi, Agullo:2012sh}, in which case many of the mathematical tools from standard quantum field theory on a curved background can be used.

The third approach to cosmological perturbation theory in LQC is motivated by the need to have a loop quantization of both the background and the perturbations.  The first approach only provides an effective framework without specifying any quantum theory, while the second quantizes the perturbations using Fock, not loop, techniques.  Now, by using the separate universe approximation it is possible to perform a loop quantization for both the background and the perturbations.  The basic idea (which will be made more precise in Sec.~\ref{s.sep}) is that, for long-wavelength perturbations, it is possible to model the perturbed cosmological space-time as a collection of homogeneous patches that evolve independently \cite{Salopek:1990jq, Wands:2000dp}.  In such a setting, the perturbations are determined by calculating the differences between the values of the `homogeneous' fields in each patch.  This picture is very helpful in the context of LQC, since it is possible to perform a standard loop quantization of a homogeneous space-time in each patch \cite{Bojowald:2006qu, WilsonEwing:2012bx}.  Then, from the resulting quantum theory, the dynamics of the perturbations can be extracted in a straightforward fashion \cite{WilsonEwing:2012bx}.  The key advantage of this approach to cosmological perturbation theory in LQC is that it provides a loop quantization of both the background and the perturbative degrees of freedom, as shall be described in Sec.~\ref{ss.sep-comments}, while the main drawback is that it is only applicable to the long-wavelength modes of the perturbations.

Once the equations of motion for cosmological perturbations in LQC are known, they can be used to calculate observables like the amplitude of the scalar perturbations, the scalar spectral index, the tensor-to-scalar ratio, and more, for any number of cosmological models.  Clearly, the predictions will depend on the specific model that is considered, but in some cases there may arise some quantum gravity effects that leave an imprint on the CMB that could be detected.  For this reason, it is important to consider as many realistic cosmological scenarios as possible within the context of LQC in order to obtain a good understanding of the types of effects that could arise in LQC.  For example, it has been found that in some cases the tensor-to-scalar ratio is strongly suppressed across the bounce due to LQC effects \cite{WilsonEwing:2012pu, Cai:2014jla}, while in others a power asymmetry is predicted in the CMB \cite{Agullo:2015aba, Gupt:unp}.

In this paper, I will start in Sec.~\ref{s.sep} by reviewing the separate universe picture and show how it can be used to derive the equations of motion for long-wavelength scalar perturbations in LQC.  Then, these equations of motion are used in Sec.~\ref{s.sc-bounce} in order to calculate how the scalar perturbations evolve across the bounce in the case where the background equation of state is constant, and the analogous calculation for tensor modes is presented in Sec.~\ref{s.r-bounce}, this time using equations of motion obtained in the effective approach to cosmological perturbation theory in LQC \cite{Cailleteau:2012fy} (this is necessary since the separate universe approach, in its current form in LQC, is only applicable to scalar modes).  An important result is that, under certain conditions, the tensor-to-scalar ratio can be significantly suppressed by quantum gravity effects during the LQC bounce.  I end with a discussion in Sec.~\ref{s.disc}.

\section{Separate Universes in LQC}
\label{s.sep}

In this section, I will first briefly review cosmological perturbation theory on a spatially flat Friedmann-Lema\^itre-Robertson-Walker (FLRW) space-time and in particular the separate universe approach developed in \cite{Salopek:1990jq, Wands:2000dp}, and then explain how it can be used in order to obtain the equations of motion for long-wavelength scalar perturbations in LQC.

\subsection{Review of the Separate Universe Picture}
\label{ss.rev-sep}

The temperature anisotropies observed in the CMB can be related to perturbations in the space-time curvature, and a convenient gauge-invariant variable that captures these perturbations is the co-moving curvature perturbation $\mR$.  For an introduction to cosmological perturbation theory, see, e.g., \cite{Mukhanov:1990me, Mukhanov:2005sc}.  In the absence of entropy perturbations, the dynamics for $\mR$ in general relativity is given by the Mukhanov-Sasaki equation (in Fourier space)
\be \label{gr-ms}
v_k'' + c_s^2 k^2 v_k - \f{z''}{z} v_k = 0, \qquad v = z \mR,
\ee
where primes denote derivatives with respect to conformal time and
\be
z = \f{a \sqrt{\rho + P}}{c_s H},
\ee
with $a$ being the scale factor, $\rho, P$ and $c_s$ being the energy density, pressure and sound speed of the matter fields respectively, and $H$ being the Hubble rate (in proper time).

For long-wavelength modes (which correspond to Fourier modes whose physical wavelength $a/k$ is much greater than the sound horizon $r_h = 1 / c_s H$), the $c_s^2 k^2$ term in \eqref{gr-ms} is negligible compared to $z''/z$, and in this limit the equation of motion can be rewritten in a simple fashion directly in terms of $\mR_k$,
\be \label{eom-Rk}
\mR_k'' + 2 \f{z'}{z} \mR_k' = 0,
\ee
with the solution
\be \label{sol}
\mR_k = A_k + B_k \int^\eta \f{d \t\eta}{z(\t\eta)^2},
\ee
where $\eta$ denotes conformal time, and $A_k$ and $B_k$ are constants of integration.

The simple form of the equation of motion \eqref{eom-Rk}, which does not contain any spatial derivative terms, motivates the separate universe picture where the dynamics of these long-wavelength modes are modeled as a collection of non-interacting universes evolving independently \cite{Wands:2000dp}.

To see this in more detail, consider splitting the spatial slices of a spatially flat FLRW space-time into a large number of co-moving `separate universe patches' that are each larger than the sound horizon.  Taking the metric of the background spatially flat FLRW space-time as
\be
ds^2 = -N(t)^2 dt^2 + a(t)^2 d \v x^2,
\ee
and assuming for simplicity the topology of the space-time to be $\mathbb{R} \times \mathbb{T}^3$, the Cartesian coordinates can be chosen so that $x_i \in [0, 1)$.  A simple discretization of the space-time into a number of separate universe patches is by splitting the space-time into $n_{tot} = 1/\ell_n^3$ cubes that each have a physical edge length of $a(t)\ell_n > r_h$.  It is convenient here to choose $\ell_n = 1 / 2m$ with $m$ a positive integer.  Therefore, in this discretization, the shortest physical wavelength that can be resolved is $2 a(t) \ell_n$ (and in terms of the co-moving coordinates $x_i$, the minimal wavelength is $2 \ell_n$ and the maximal wavenumber is $k_{max} = 1 / 2 \ell_n$).

Given this choice of separate universe patches, the average co-moving curvature perturbation in a particular patch $n$ is
\begin{align}
\mR_n &= \f{1}{\ell_n^3} \int_n d^3 \v x \, \mR(\v x) \\
&= \f{1}{(2\pi)^{3/2} \ell_n^3} \int_n d^3 \v x \int_0^{k_{max}} d^3 \v k \, \mR_{\v k} \, e^{i \v k \cdot \v x},
\end{align}
where the upper limit for each of the $k_i$ is $k_{max} = 1 / 2 \ell_n$.  Note that the Fourier modes with at least one $k_i > k_{max}$ do not contribute to the value of $\mR_n$ since $\int_n d^3 \v x \exp(i \v k \cdot \v x) = 0$ for these modes.  Indeed, in the separate universe picture discrete Fourier transformations relate the $\mR_n$ to only the long-wavelength (i.e., all $k_i \le k_{max}$) modes of $\mR_{\v k}$, and vice versa.

The key point here is that the evolution of $\mR_n$ is completely determined by the evolution of the long-wavelength modes $\mR_{\v k}$ with all $k_i \le k_{max}$, and the equation of motion for these modes is \eqref{eom-Rk}.  From this, it follows that
\be \label{sep-cl}
\mR_n'' + 2 \f{z'}{z} \mR_n' = 0.
\ee
Clearly, the evolution of $\mR_n$ in one patch is entirely determined by $\mR_n$ itself and of the background FLRW space-time.  In other words, there are no derivative or interaction terms between different super-horizon patches, and in this sense each patch can be thought of as a separate universe that evolves independently, as anticipated earlier.  Thus, for super-horizon modes, the co-moving curvature perturbations can be studied by first discretizing the space-time into separate universe patches that are larger than the sound horizon, and then calculating the evolution of the perturbations in each patch, neglecting any interactions between neighbouring patches.

It is important to emphasize that this result follows from the simple fact that for super-horizon Fourier modes the $c_s^2 k^2$ term in \eqref{gr-ms} is subdominant compared to $z''/z$.  So long as this condition holds then the separate universe approximation can be used.

As an aside, note that in the literature the separate universe approximation is often associated to the result that in an expanding universe the second term in \eqref{sol} decays rapidly, and hence the super-horizon modes of $\mR_k$ freeze and become constant.  However, as explained above, the reason the separate universe approximation holds for super-horizon modes is not that the co-moving curvature perturbation freezes, but rather that the $c_s^2 k^2$ term is subdominant compared to $z''/z$ in \eqref{gr-ms}.  Therefore, the separate universe approximation can also be used in contracting space-times where the second term in \eqref{sol} is instead a growing mode and in fact becomes the dominant term.  The validity of the separate universe approximation for long-wavelength modes in both contracting and expanding universes (as well as during a bounce) will be important in what follows.

Finally, while the separate universe picture can be used in both expanding and contracting universes, it does have some important limitations.  Clearly, it is only applicable for long-wavelength perturbations.  For short-wavelength perturbations (i.e., $a/k < r_h$), derivative terms become important and the separate universe framework cannot be used as interactions between patches would be important.  Another limitation is that here it has been assumed that there are no entropy perturbations; this is necessary since entropy perturbations appear as a source in the equation of motion for $\mR$ and can significantly affect its dynamics.  For this reason, the results obtained in this paper only concern the LQC dynamics of long-wavelength perturbations in the absence of entropy perturbations.

\subsection{General Relativity}
\label{ss.sep-bg}

The ultimate aim here is to use the separate universe approach in the context of LQC and obtain a loop quantization of both the background and the perturbations.  In order to do this, it is necessary to treat each (non-interacting) separate universe patch as a spatially flat FLRW space-time, from which both the background and perturbative degrees of freedom can be extracted.  Furthermore, given the Friedmann equations it is possible to derive the equations of motion for the long-wavelength scalar perturbations.  In this section, before going on to the case of LQC, I will explain how this can be done for general relativity minimally coupled to a scalar field $\phi$ with some potential $V(\phi)$.

Including linear scalar perturbations on a spatially flat FLRW space-time, in the longitudinal gauge and for conformal time, the line element is
\be
ds^2 = a(\eta)^2 \Big( -[1 + 2 \psi(\v x, \eta)] d\eta^2 + [1 - 2 \psi(\v x, \eta)] d\v x^2 \Big),
\ee
where it is understood that $|\psi| \ll 1$ since the perturbations are assumed to be small.  This line element can be rewritten as
\be
ds^2 = \t a(\v x, \eta)^2 \Big( -[1 + 4 \psi(\v x, \eta)] d\eta^2 + d\v x^2 \Big),
\ee
where $\t a = a (1 - \psi)$ and second order terms in the perturbation $\psi$ have been dropped.  Note that to linear order in perturbation theory (which is the truncation considered in this paper), $\t a = a (1 - \psi) \Leftrightarrow a = \t a (1 + \psi)$.

Recall that the topology of the space-time has been assumed to be $\mathbb{R} \times \mathbb{T}^3$.  An important point is that since the zero Fourier mode of any perturbation can be absorbed into the background, it is natural to require that $\int_{\mathbb{T}^3} \psi = 0$.

It is at this point that the separate universe approximation is used.  After dividing the space-time into $n_{tot}$ super-horizon patches that are each assumed to be homogeneous, the metric in any patch $n$ is given by
\be
ds_n^2 = \t a_n(\eta)^2 \Big( -[1 + 4 \psi_n(\eta)] d\eta^2 + d\v x^2 \Big),
\ee
which is exactly the metric for a spatially flat FLRW space-time, although with the unusual lapse $N_n = \t a_n (1 + 2 \psi_n)$.  In the discretized setting, the requirement that $\int_{\mathbb{T}^3} \psi = 0$ becomes $\sum_n \psi_n = 0$, and from this follow the useful relations
\be
a = \sum_n \f{\t a_n}{n_{tot}}, \qquad \psi_n = \f{a - \t a_n}{\t a_n}.
\ee
An important point is that with these relations, every term appearing in the metric can be understood as a function of the $\t a_n$.  In particular, the lapse can be rewritten as
\be \label{lapse}
N_n = 2 \sum_m \f{\t a_m}{n_{tot}} - \t a_n = 2 a - \t a_n.
\ee
With these relations, it is possible to rewrite the line element for each patch one last time as
\be \label{metric-long}
ds_n^2 = - N_n^2 d\eta^2 + \t a_n^2 d\v x^2,
\ee
with $N_n$ given in \eqref{lapse}, which as noted above is only a function of the $\t a_n$.  Here $\eta$ is a time coordinate which globally defines the equal time hypersurfaces in the entire space-time.

Similarly, the scalar field $\phi$ which in the continuum is split as $\phi = \b\phi + \de\phi$, becomes $\phi_n$ and the background and perturbation split is recovered through $\b\phi = \sum_n \phi_n / n_{tot}$ and $\de\phi_n = \phi_n - \b\phi$.

From this discussion, it is clear that the dynamics of both the background and the long-wavelength perturbations can be determined once the dynamics of the $\t a_n$ and $\phi_n$ are known.  Since every patch is homogeneous, and in the separate universe picture interactions between patches are ignored, it is now a straightforward procedure to derive the dynamics for $\t a_n$ and $\phi_n$ from the Friedmann equations.

In each patch, since the metric is that of a flat FLRW space-time and interactions are neglected, the usual Friedmann equations hold.  To be specific, for a lapse $N_n$, in each patch the equations of motion are the Friedmann equation
\be \label{cl-fr1}
\left( \f{\t a_n'}{\t a_n} \right)^2
= \f{8 \pi G}{3} \left[ \f{(\phi_n')^2}{2} + N_n^2 \, V(\phi_n) \right],
\ee
the Raychaudhuri equation
\be \label{cl-fr2}
\left( \f{\t a_n'}{\t a_n} \right)' = - 4 \pi G (\phi_n')^2 + \f{N_n' \t a_n'}{N_n \t a_n},
\ee
and the continuity equation
\be \label{cl-cont}
\left[\f{(\phi_n')^2}{2 N_n^2} + V(\phi_n) \right]' + 3 \f{\t a_n' (\phi_n')^2}{\t a_n N_n^2} = 0.
\ee
Primes denote derivatives with respect to the time coordinate $\eta$, and recall that the energy density and pressure of the scalar field in each patch are $\rho_n = (\phi_n')^2 / 2 N_n^2 + V(\phi_n)$ and $P_n = (\phi_n')^2 / 2 N_n^2 - V(\phi_n)$ respectively.

Then, by expanding to linear order in perturbations \eqref{cl-fr1}, \eqref{cl-fr2} and \eqref{cl-cont} using the relations $\t a_n = a (1 - \psi_n)$, $N_n = a (1 + \psi_n)$ and $\phi_n = \b\phi + \de\phi_n$, it is possible to recover the equations of motion for the background and for long-wavelength perturbations to linear order.  It is immediate that for the background the Friedmann and Raychaudhuri equations in conformal time are recovered,
\be
\mH^2 := \left( \f{a'}{a} \right)^2 = \f{8 \pi G}{3} \left[ \f{(\b\phi')^2}{2} + a^2 \, V(\b\phi) \right],
\ee
\be
\f{a''}{a} - 2 \mH^2 = - 4 \pi G (\b\phi')^2,
\ee
and the continuity equation gives precisely the Klein-Gordon equation for $\b\phi$,
\be
\b\phi'' + 2 \mH \b\phi' + a^2 V_\phi(\b\phi) = 0,
\ee
where $V_\phi(\phi) := dV(\phi)/d\phi$.

The equations of motion for the perturbations derived from this separate universe approximation are, from the Friedmann equation,
\be
3 \mH \psi_n' = - 4 \pi G \left[ 2 a^2 \, V \psi_n + \b\phi' \de\phi_n' + a^2 \, V_\phi \de\phi_n \right],
\ee
where it is understood that $V = V(\b\phi)$ and $V_\phi = V_\phi(\b\phi)$, and from the Raychaudhuri equation,
\be
\psi_n'' = 8 \pi G \b\phi' \de\phi_n',
\ee
while the continuity equation gives
\be \label{pert-phi}
\de\phi_n'' +  2 \mH \de\phi_n' + a^2 V_{\phi\phi} \de\phi_n = 4 \b\phi' \psi_n' - 2 a^2 \, V_\phi \psi_n,
\ee
where $V_{\phi\phi} := d^2 V(\b\phi) / d\phi^2$.

It is straightforward to check that these equations, while not in the standard textbook form, are nonetheless exactly equivalent to the standard equations of motion ---in the longitudinal gauge and for long-wavelength modes--- of cosmological perturbation theory for a flat FLRW space-time with a scalar field.

This shows how, given the Friedmann equations for a flat FLRW space-time, the separate universe approach provides a clear path to derive the equations of motion for long-wavelength perturbations.  The next step is to repeat this procedure in the context of LQC where the Friedmann equations are modified by quantum gravity effects and hence the dynamics of the perturbations may also be modified.

However, before applying the separate universe picture to LQC, a few comments are in order.  As mentioned at the end of Sec.~\ref{ss.rev-sep}, the separate universe picture is applicable to the study of long-wavelength perturbations in the absence of entropy perturbations.  For the extension presented here that extracts the equations of motion for the long-wavelength scalar perturbations from the Friedmann equations, there are a few additional restrictions that arise.  The procedure requires the metric to be in the form \eqref{metric-long}, which is only possible for scalar perturbations, and then only in the longitudinal gauge.  It may be possible to extend this treatment to include vector and tensor modes, or to do it in a gauge-invariant fashion, but this would require using a more general metric in each patch than the flat FLRW one, perhaps a Bianchi space-time or some other more general homogeneous space-time.  However, this remains an open problem at this time.  Also, note that in general the metric in the longitudinal gauge has the form $ds^2 = -a^2(1+2\varphi) d\eta^2 + a^2(1-2\psi) d\v x^2$, but since a scalar field has no anisotropic stress, it is possible to set $\varphi=\psi$.  This simplification is also necessary in order to be able to put the metric in the form \eqref{metric-long}.  Therefore, the results presented here in Sec.~\ref{ss.sep-bg} can only be obtained for scalar perturbations in the longitudinal gauge, and then only if the matter field has no anisotropic stress.

\subsection{Loop Quantum Cosmology}
\label{ss.sep-lqc}

In loop quantum cosmology, the non-perturbative quantization techniques of loop quantum gravity are applied to cosmological space-times where there are only a finite number of degrees of freedom.  This approach has been very successful, but it has been difficult to extend these results in order to obtain a full loop quantization of a homogeneous cosmological space-time with small perturbations.  However, as shall be shown here, the separate universe picture provides a framework to do exactly this, at least for long-wavelength perturbations.  This section builds on earlier related work in \cite{Artymowski:2008sc, WilsonEwing:2011es, WilsonEwing:2012bx}.

In order to see how this can be done, it is important to recall some relevant results of LQC for the flat FLRW space-time.  In particular, while the quantum theory has been studied in considerable detail, what is most important here are the effective dynamics of LQC.  Since the observables of interest in homogeneous space-times are global quantities (for example, the volume $V = a^3$ or the value of the scalar field $\phi$), they correspond to heavy degrees of freedom in the quantum mechanical sense that the quantum fluctuations in these degrees of freedom do not grow significantly if the quantum fluctuations are initially small \cite{Rovelli:2013zaa}.  Therefore, assuming that the wave function is initially sharply peaked, it is enough to study the dynamics of the expectation values of the observables $\mO$ of interest, and the approximation $\bra \mO^n \ket \approx \bra \mO \ket^n$ will be preserved in time for these heavy degrees of freedom.

Then, for these sharply peaked states the dynamics of the expectation value in LQC of the scalar field and the scale factor are given by the LQC effective Friedmann equations \cite{Ashtekar:2006wn, Taveras:2008ke}
\be \label{lqc-fr1}
\left( \f{a'}{a} \right)^2
= \f{8 \pi G N^2}{3} \rho \left( 1 - \f{\rho}{\rho_c} \right),
\ee
\be \label{lqc-fr2}
\left( \f{a'}{a} \right)' = - 4 \pi G N^2 (\rho + P) \left( 1 - \f{2 \rho}{\rho_c} \right) + \f{N' a'}{N a},
\ee
where $\rho_c \sim \rho_{\rm Pl}$ is the critical energy density of LQC.  Note that the continuity equation
\be \label{lqc-cont}
\rho' + 3 \left( \f{a'}{a} \right) (\rho + P) = 0
\ee
is not modified by any quantum gravity effects.  While these equations are usually presented in terms of proper time (i.e., $N=1$), it is easy to generalize their form for any choice of lapse, as has been done here.  Using the separate universe approximation, the LQC effective equations of motion for scalar perturbations can be derived from these equations.

Before doing this, it is important to discuss the framework that is being considered here in further detail.  The equations being derived here are effective equations for the perturbations: a more careful treatment (which has been developed but is unwieldy for explicit calculations, see \cite{WilsonEwing:2012bx}) is necessary in order to study any effects due to quantum fluctuations.  Therefore, these effective equations can be used to calculate the evolution of those perturbations that can be treated classically before the bounce.  This is of interest in a number of contexts, and in particular in alternatives to inflation like the matter bounce scenario \cite{Brandenberger:2012zb} and the ekpyrotic universe \cite{Lehners:2008vx}; in both cases, although the perturbations originate from quantum fluctuations, the perturbations are squeezed in the contracting space-time and can be treated classically by the time the bounce occurs.  (For a description of this quantum to classical transition in cosmological perturbation theory through squeezing, see, e.g., \cite{Polarski:1995jg, Battarra:2013cha}.)  On the other hand, this framework would need to be extended in order to handle cosmological scenarios like inflation, where quantum fluctuations play an important role during the bounce.  (In addition, for inflation the wavelengths of interest are typically trans-Planckian and then the separate universe approximation fails.)  So, it is important to keep in mind that the effective equations derived in this section can only be used for perturbations that can be treated in a classical fashion before the bounce, and that to include the effect of quantum fluctuations it will be necessary to instead use the full quantum theory described in Sec.~\ref{ss.sep-comments}.

Now, by following exactly the same procedure as in Sec.~\ref{ss.sep-bg}, it is possible to derive the equations of motion for long-wavelength cosmological perturbations in LQC.  By considering a collection of $n_{tot}$ homogeneous patches labeled by $n$, each with the Friedmann equations \eqref{lqc-fr1} and \eqref{lqc-fr2}, and with a scale factor $\t a_n = a(1 - \psi_n)$, value of the scalar field $\phi_n = \b\phi + \de\phi_n$ and lapse $N_n = a(1 + \psi_n)$, and recalling that $\rho_n = (\phi_n')^2/2 N_n^2 + V(\phi_n)$ and $\rho_n + P_n = (\phi_n')^2/N_n^2$, it is easy to verify that the correct LQC effective Friedmann equations for the lapse $N=a$ are recovered for the background,
\be
\mH^2 = \f{8 \pi G}{3} a^2 \b\rho \Big( 1 - \f{\b\rho}{\rho_c} \Big),
\ee
\be
\mH' - \mH^2 = -4 \pi G (\b\phi')^2 \Omega,
\ee
where $\b\rho = (\b\phi')^2 / 2 a^2 + V(\b\phi)$ and $\Omega = 1 - 2 \b\rho / \rho_c$.

With a little more work, it can be checked that the equations of motion for the perturbations are, from \eqref{lqc-fr1}
\be \label{lqc-pert1}
\mH \psi_n' + \mH^2 \psi_n = - \f{4 \pi G}{3} a^2 \Omega \, \de\rho_n,
\ee
and from \eqref{lqc-fr2},
\be \label{lqc-pert2}
\psi_n'' = 8 \pi G \, \Omega \, \b\phi' \de\phi_n' - \f{8 \pi G}{\rho_c} (\b\phi')^2 \de\rho_n,
\ee
with
\be \label{del-rho}
\de\rho_n = \f{1}{a^2} \Big( \b\phi' \de\phi_n' - (\b\phi')^2 \psi_n + a^2 V_\phi \de\phi_n \Big).
\ee
As before, the continuity equation can be expanded in order to obtain the equation of motion for $\de\phi_n$.  Since the continuity equation has the same form in LQC as in classical general relativity, the equation of motion for $\de\phi_n$ is \eqref{pert-phi}, just as in the classical theory.  (This result follows from the fact that at no point does the derivation of \eqref{pert-phi} require the use of the Friedmann equations.)

The diffeomorphism constraint for long-wavelength perturbations has the form, in LQC, of
\be \label{lqc-diff}
D = \psi_n' + \mH \psi_n - 4 \pi G \, \Omega \, \b\phi' \de\phi_n' = 0.
\ee
This result can be derived in the LQC separate universe framework by noticing that $\Omega D' + 2 \mH \Omega D - (6 \mH (\b\phi')^2 / \rho_c a^2) D = 0$, which can be shown to be zero from \eqref{pert-phi} and \eqref{lqc-pert2}.  Then, the only solution to this differential equation for $D$ being first order in perturbation theory is $D=0$.

Another useful relation is obtained by combining \eqref{lqc-pert1} and \eqref{lqc-diff},
\be
a^2 \de\rho_n + 3 \mH \b\phi' \de\phi_n = 0.
\ee

From these equations, using the definition that in the longitudinal gauge $v_n = z \psi_n + a \de\phi_n$, after some algebra it can be shown that
\be
v_n'' - \f{z''}{z} v_n = 0,
\ee
just as in the classical theory, and it then immediately follows that, again just as in the classical theory,
\be \label{sep-lqc}
\mR_n'' + 2 \f{z'}{z} \mR_n' = 0.
\ee
The key difference here is that $z$ depends on the background variables which follow the equations of motion of LQC, not those of classical general relativity.

It is important to remember that \eqref{sep-lqc} can only be used for long-wavelength modes.  In particular, near the bounce where quantum gravity effects are important, long-wavelength modes correspond to Fourier modes whose physical wavelength is larger than $\lp$, and therefore these equations cannot be used to study trans-Planckian modes.

However, so long as one is only interested in Fourier modes whose wavelength remains larger than the Planck length at all times, \eqref{sep-lqc} can be used in order to calculate how these modes evolve across the LQC bounce, assuming their form is known in the contracting pre-bounce branch.

\subsection{Some Comments on the Quantum Theory}
\label{ss.sep-comments}

In the previous section I derived the effective equations for long-wavelength scalar perturbations from the effective Friedmann equations of LQC.  It is important to keep in mind that these effective equations come from a complete loop quantization of the background and perturbations in the separate universe approximation.  Now, I will give a streamlined presentation of this loop quantization; for further details see \cite{WilsonEwing:2012bx}.

Motivated by the separate universe approximation, the idea is to consider a collection of homogeneous patches, each of which is treated as a spatially flat FLRW space-time.  Then, each patch can be quantized following the standard procedure of LQC.

Therefore, the total kinematical Hilbert space is a tensor product of the kinematical Hilbert spaces for each patch, $H_{\rm tot} = \otimes_n H_n$, where the Hilbert space for each patch $H_n$ is the standard kinematical Hilbert space for isotropic LQC defined in \cite{Ashtekar:2003hd}.  Then, the dynamics are generated by a total Hamiltonian constraint operator,
\be
\wh{\mC_H} = \sum_n \wh{N_n} \wh{\mH_n},
\ee
where $\wh{\mH_n}$ in each patch is the usual scalar constraint operator for spatially flat FLRW space-times given in, e.g., \cite{Ashtekar:2006wn, Ashtekar:2007em} and $\wh{N_n}$ is an operator constructed out of operators corresponding to the scale factors of each patch $\wh{\t a_n}$ (or $\wh{\sqrt{p_n}}$ in terms of the densitized triad variables $p = a^2$) according to the first equality in \eqref{lapse}.  Finally, the physical Hilbert space is composed of the states annihilated by all of the Hamiltonian constraint operators $\wh{\mH_n}$.  This completes the full definition of the quantum theory.

Then, from this quantum theory, for sharply peaked states it is possible to extract the effective equations in each patch which are precisely \eqref{lqc-fr1}, \eqref{lqc-fr2} and \eqref{lqc-cont} \cite{Ashtekar:2006wn, Taveras:2008ke, Rovelli:2013zaa}, from which the LQC effective equations of motion for the long-wavelength scalar perturbations \eqref{lqc-pert1}, \eqref{lqc-pert2}, \eqref{pert-phi} and then \eqref{sep-lqc} follow.

I will end this section with a few technical comments about the quantum theory for the interested reader; for a more complete discussion see \cite{WilsonEwing:2012bx}.

First, since $N_n$ depends on the $\t a_m$ of all patches, the dynamics of the theory is non-local.  It is natural to worry that this may lead to anomalies in the quantum theory, but since $N_n$ multiplies the scalar constraint, all potential non-local anomalies in the constraint algebra vanish in the physical Hilbert space as can readily be checked by a short calculation.

Second, while this quantum system is well-defined in general, for a state in the Hilbert space to correspond to a cosmological space-time with small perturbations, it is necessary to consider states where all patches are ``similar'' to each other.  A natural way to implement this is through the requirement that, for all relevant operators (the volume $V=a^3$, $\phi$ and appropriate expressions of their conjugate momenta), the difference between expectation values in any two patches must be much smaller than its average expectation value in all patches, i.e.,
\be \label{small-perts}
\bra \wh{\mO_{n_1}} - \wh{\mO_{n_2}} \ket \ll \sum_m \f{\bra \wh{\mO_m} \ket}{n_{tot}}.
\ee
Interestingly, a necessary condition for this inequality to hold is for the physical volume of each patch to be larger than the Planck volume \cite{WilsonEwing:2012bx}.  Thus, if the volume of any patch becomes smaller than the Planck volume due to the evolution from the Hamiltonian constraint operator, the separate universe approximation breaks down, and while the quantum theory remains well-defined, the physical system no longer provides an approximation to a homogeneous cosmological space-time with small (long-wavelength) perturbations.

A third point concerns the question of gauge-fixing before quantization: the longitudinal gauge was imposed before quantization, and in fact determined the form the lapse should have.  While it would be nice to have a gauge-invariant treatment, it is not immediately obvious how to extend the results obtained here to a gauge-invariant framework.  In order to check the validity of gauge-fixing before quantizing, the key question is whether the lapse in the longitudinal gauge (and in the absence of anisotropic stress) continues to have the form $N = a(1+\psi)$, which is assumed in this treatment.  Clearly, this question cannot be properly addressed within the context of this framework.  However, in the approach to cosmological perturbation theory in LQC based on consistent deformations of the constraint algebra in the effective theory (one of the two alternatives to the separate universe approach), it is possible to derive the effective equations without choosing any gauge, and from this it can be checked that the longitudinal gauge is a valid gauge in that framework and that in that gauge the relation $N = a(1+\psi)$ always holds, even at the bounce point \cite{Cailleteau:2011kr}.  Furthermore, the effective equations of motion for the perturbations derived in this framework, when set in the longitudinal gauge and after taking the long-wavelength limit, exactly agree with the results found in the separate universe approach.  This provides some justification that the longitudinal gauge remains valid at all times, including at the bounce point, and that the separate universe procedure described here provides a good approximation to the full dynamics of long-wavelength scalar perturbations.

Finally, it is possible to include interactions between neighbouring patches in this framework \cite{WilsonEwing:2012bx}, and one might hope that this could lead to a treatment that could handle small-wavelength perturbations during the LQC bounce in addition to long-wavelength modes.  However, since these interactions are only relevant for wavelengths that are comparable to or smaller than the curvature radius ---which is of the order of the Planck length near the bounce--- these interaction terms are only important for trans-Planckian modes.  Since these are precisely the modes for which the volume of each patch is of the order of the Planck volume, in this case the condition \eqref{small-perts} necessarily fails for some observable and therefore the system does not correspond to a homogeneous cosmology with small perturbations.  For this reason, the `patchwork' framework described here (without or without interactions) can only be applied to perturbations whose wavelength is large compared to $\lp$, and these are precisely the modes for which the interaction terms are negligible near the bounce.  Therefore, there is no advantage to including any interactions in a framework like this one where the aim is describing LQC effects on cosmological perturbations.

\section{The Evolution of Scalar Perturbations Across the Bounce}
\label{s.sc-bounce}

In this section, using the results of the separate universe approach to cosmological perturbation theory in LQC, it will be shown how to calculate the evolution of the scalar perturbations across the LQC bounce.  For simplicity, I will assume that the equation of state $\om$ (relating the pressure $P$ and the energy density $\rho$ of the matter fields via $P = \om \rho$) is constant.  Nonetheless, it will be clear how these calculations can be generalized to allow for a dynamical equation of state.  I will also assume here that the perturbations can be treated classically before the bounce.

The goal here is to determine the form of the co-moving curvature perturbation after the bounce, assuming that it is known in the contracting branch at times before any quantum gravity effects become important (as is the case, for example, for matter bounce \cite{Brandenberger:2012zb} and ekpyrotic cosmologies \cite{Lehners:2008vx}).  The scale factor of a contracting spatially flat FLRW space-time in general relativity is given by
\be \label{class-a}
a(t) = a_o |t|^{2/(3+3\om)}, \qquad -\infty < t < 0.
\ee
Then, since in general relativity when $\om$ is constant $z \propto a$, the solution to the equation of motion \eqref{sol} for $\mR_k$ in the long-wavelength limit is
\be
\mR_k^- = A_k + B_k \int \f{d\eta}{a(t)^2},
\ee
where $B_k$ has been redefined in order to absorb some numerical prefactors and the superscript `$-$' is to denote that this is the form of $\mR_k$ well before the bounce.  Then, through the relation $a \, d\eta = dt$, the integral can easily be evaluated, giving
\be \label{classR}
\mR_k^- = A_k + B_k |t|^{(\om-1)/(1+\om)},
\ee
where once again $B_k$ has been redefined in order to absorb numerical prefactors.

Now the task is to calculate the evolution of $\mR_k$ across the LQC bounce, assuming that $A_k$ and $B_k$ are known.  This will be done by solving for the evolution of $\mR_k$ when LQC effects are included which I will denote by $\mR_k^q$ --- this result will be valid at all times for long-wavelength modes.  In particular, in the limit $\lim_{t\to-\infty} \mR_k^q$ (which corresponds to the contracting pre-bounce universe where quantum gravity effects are negligible) the solution \eqref{classR} must be recovered.  Imposing that $\mR_k^-$ and $\mR_k^q$ agree at early times will entirely determine the form of $\mR_k^q$, and then the form of the co-moving curvature perturbations after the bounce will simply be given by $\lim_{t\to\infty} \mR_k^q$.

Finally, note that $A_k$ and $B_k$ will typically depend on the Fourier number $k$ and, if these modes are scale-invariant, will go as $k^{-3/2}$.  However, the analysis that follows is general and does not depend on the scale-dependence of $A_k$ and $B_k$ on $k$.

\subsection{The LQC Bounce}
\label{ss.bg}

For a constant equation of state $\om$, the LQC Friedmann equations can be solved exactly in terms of the proper time $t$ (i.e., for $N=1$),
\be \label{lqc-a}
a(t) = \Big( 6 \pi G \rho_c (1+\om)^2 t^2 + 1 \Big)^{1/(3+3\om)},
\ee
where the constants of integration are chosen such that the bounce occurs at $t=0$, and such that $a(t=0)=1$.  Given these choices, the energy density of the matter field is given by
\be
\rho(t) = \f{\rho_c}{a(t)^{3(1+\om)}}.
\ee

The solution for $\mR_k$ in the long-wavelength limit when LQC effects are included, using the fact that $z \propto a / \sqrt{1 - \rho / \rho_c}$ in LQC for constant $\om$, is
\be
\mR_k = C_k + D_k \int \f{d\eta}{a(t)^2} \left(1 - \f{\rho}{\rho_c} \right),
\ee
where $D_k$ has been redefined to absorb some numerical prefactors.  As before, the integral can be expressed in terms of proper time $t$ by using the relation $a \, d\eta = dt$, which gives
\be
\mR_k = C_k + D_k \int dt \f{\alpha t^2}{a(t)^{3(2+\om)}},
\ee
where $\alpha = 6 \pi G \rho_c (1+\om)^2$, and this integral can be evaluated to give
\be \label{lqcR}
\mR_k^q = C_k + D_k \, \f{\alpha t^3}{3} \,
{}_2F_1 \! \left( \tf{3}{2}, \tf{2+\om}{1+\om}; \tf{5}{2}; -\alpha t^2 \right),
\ee
where ${}_2F_1$ are the hypergeometric functions.

As explained above, for $t \ll -1/6 \pi G \rho_c$ ---i.e., in the contracting branch before any quantum gravity effects become important--- the solution \eqref{lqcR} must match \eqref{classR}.  This will uniquely determine $C_k$ and $D_k$ in terms of $A_k$ and $B_k$.  To do this, it is enough to take the large $-t$ limit in \eqref{lqcR} which gives \cite{Abramowitz-Stegun}
\begin{align} \label{expR}
\mR_k^q \to & \, \,
C_k - D_k \f{\sqrt\pi \, \Gamma(\tf{2+\om}{1+\om}-\tf{3}{2})}{4 \sqrt\alpha \, \Gamma(\tf{2+\om}{1+\om})} \nn \\ &
- \f{D_k}{\alpha^{1/(1+\om)}} \left(\f{1+\om}{\om-1} \right) |t|^{(\om-1)/(1+\om)}.
\end{align}
A technical point here is that the terms written in this equation are actually not necessarily the leading order terms to $\mR_k^q$ as $t\to-\infty$, instead they are the terms that correspond to the classical solution.  In some cases (for $\om < -1/3$), there are LQC corrections which are subleading to the constant term, but go to zero slower than $|t|^{(\om-1)/(1+\om)}$ as $t\to-\infty$.  However, since these terms are quantum gravity corrections to the classical solution, they are irrelevant for matching the LQC solution to \eqref{classR}.

Given \eqref{expR}, it is easy to check that requiring that $\mR_k^- = \mR_k^q$ in the $t \ll -1 / 6 \pi G \rho_c$ limit determines $C_k$ and $D_k$ to be
\be
C_k = A_k - B_k \, \alpha^{\tf{1-\om}{2(1+\om)}} \left(\f{\om-1}{1+\om}\right) \f{\sqrt\pi \, \Gamma(\tf{2+\om}{1+\om}-\tf{3}{2})}{4 \, \Gamma(\tf{2+\om}{1+\om})},
\ee
\be
D_k = B_k \, \alpha^{1/(1+\om)} \left( \f{\om-1}{1+\om} \right).
\ee

Now in order to determine the form $\mR_k$ after the bounce, all that is required is to consider the limit of $t \gg 1 / 6 \pi G \rho_c$ in \eqref{lqcR}.  The dominant term is constant in time,
\be
\mR_k^+ = C_k + D_k \f{\sqrt\pi \, \Gamma(\tf{2+\om}{1+\om}-\tf{3}{2})}{4 \sqrt\alpha \, \Gamma(\tf{2+\om}{1+\om})},
\ee
where the `$+$' superscript denotes that this corresponds to the value of $\mR_k$ after the bounce, and this result can be expressed in terms of the $A_k$ and $B_k$ of \eqref{classR}, giving
\be
\mR_k^+ = A_k - B_k \, \alpha^{\tf{1-\om}{2(1+\om)}} \left(\f{\om-1}{1+\om}\right) \f{\sqrt\pi \, \Gamma(\tf{2+\om}{1+\om}-\tf{3}{2})}{2 \, \Gamma(\tf{2+\om}{1+\om})}.
\ee

This result can be further simplified via considerations from the contracting branch.  In the contracting branch, in the regime where quantum gravity effects are negligible, the solution \eqref{classR} contains two terms, one constant and the other varying with time.  If $-1 < \om < 1$, then the time-dependent term is a growing mode and typically quickly dominates the constant mode --- in this case the constant mode is negligible and $A_k$ can be set to zero.  On the other hand, if the contraction is ekpyrotic and $\om > 1$, then the time-dependent term is a decaying mode as $t \to 0$, in which case the constant term dominates and $B_k$ is negligible.

Therefore, for $-1 < \om < 1$,
\be
\mR_k^+ = B_k \, \alpha^{\tf{1-\om}{2(1+\om)}} \left(\f{1-\om}{1+\om}\right) \f{\sqrt\pi \, \Gamma(\tf{2+\om}{1+\om}-\tf{3}{2})}{2 \, \Gamma(\tf{2+\om}{1+\om})},
\ee
and for $\om > 1$,
\be
\mR_k^+ = A_k.
\ee

Two comments are in order here.  First, the case of $\om=1$ is a limiting case where the analysis performed above is not immediately applicable.  However, it is strightforward to extend these results for $\om=1$ by following the same procedure with the result that, assuming that in the contracting branch $\mR_k$ is given by
\be \label{om1-bef}
\mR_k^- = A_k + B_k \ln |t|,
\ee
(note that the logarithmic term decays as $t$ approaches zero, the divergence in the logarithm at $t=0$ is resolved once quantum gravity effects are included) then in the expanding branch, after the bounce,
\be \label{om1-aft}
\mR_k^+ = A_k - B_k \ln |t|,
\ee
with the logarithmic term now corresponding to a growing term as $t$ increases.

The second comment concerns space-times where the matter field has an equation of state $\om \ge 1$.  In the expanding branch, the long-wavelength modes actually contain a growing mode which in the limit of $t \to \infty$ will dominate the constant mode.  However, its amplitude will initially be extremely small (since its amplitude is determined only by the amplitude of the decaying mode in the contracting branch).  Therefore, for quite some time after the bounce, the constant mode will dominate the growing mode, and so long as the transition from the ekpyrotic era around the bounce to standard cosmology occurs soon after the bounce, the constant mode is the dominant contribution to $\mR_k^+$.  However, if one is interested in a scenario where the equation of state remains $\om \ge 1$ for a long time after the bounce, then it may be necessary to include the growing mode in the analysis.

\section{The Tensor-to-Scalar Ratio Across the Bounce}
\label{s.r-bounce}

After calculating how the scalar perturbations evolve across the LQC bounce, the natural next step is to consider the same question for tensor perturbations.  As shall be shown in this section, a similar procedure can be followed for the long-wavelength Fourier modes of the tensor perturbations $h_k$ in order to calculate, given the form of the tensor modes in the classical pre-bounce era, their evolution across the bounce.

Since the equation of motion for long-wavelength $h_k$ in the regime where quantum gravity effects are negligible is
\be \label{class-h}
h_k'' + \f{2 a'}{a} h_k' = 0,
\ee
the solution is
\be
h_k = \t A_k + \t B_k \int \f{d\eta}{a(t)^2} = \t A_k + \t B_k \int \f{dt}{a(t)^3},
\ee
and, for the case when the equation of state $\om$ for the matter field is constant and the scale factor is given by \eqref{class-a}, the integral can be evaluated,
\be \label{class-sol-h}
h_k^- = \t A_k + \t B_k |t|^{(\om-1)/(1+\om)},
\ee
where $\t B_k$ has been redefined in order to absorb some numerical prefactors and the superscript `$-$' denotes that this corresponds to the pre-bounce form of $h_k$.

Choosing some appropriate initial conditions in the contracting space-time before the bounce will specify $\t A_k$ and $\t B_k$, and the task here is to determine the post-bounce form of $h_k$, for the pre-bounce form of $h_k$ given in \eqref{class-sol-h}.

\subsection{Tensor Perturbations in LQC}
\label{ss.tensor}

To do this, it is necessary to know how $h_k$ evolves when LQC effects are important.  While the separate universe approach to cosmological perturbation theory as described in Sec.~\ref{s.sep} can only be applied directly to scalar modes, there do exist other approaches to cosmological perturbation theory in LQC that can more easily handle tensor perturbations.  In particular, as mentioned in the Introduction there is an effective Hamiltonian approach which is based on including LQC corrections in the effective Hamiltonian and diffeomorphism constraints that correspond to holonomy or inverse triad corrections, and requiring that the resulting constraint algebra be anomaly-free \cite{Bojowald:2008jv}.

Interestingly, when this approach to cosmological perturbation theory is used in order to determine the effect of holonomy corrections to scalar perturbations, exactly the same equations of motion are obtained as those presented in Sec.~\ref{ss.sep-lqc} that follow from the separate universe approach \cite{Cailleteau:2011kr}.  The fact that these two approaches give the same results provides evidence for the robustness of the equations of motion for scalar perturbations in LQC presented in this paper.

Furthermore, this robustness also lends weight to the results of the effective Hamiltonian approach when applied to tensor perturbations, where the equation of motion for long-wavelength Fourier modes of $h_k$ is \cite{Cailleteau:2012fy}
\be \label{lqc-h}
h'' + \f{2 z_T'}{z_T} h' = 0, \qquad z_T = \f{a(t)}{\sqrt{1 - 2 \rho / \rho_c}},
\ee
for the case where holonomy corrections have been included.  This is the equation I will use in order to study the evolution of tensor perturbations across the LQC bounce.

The solution of this equation of motion is given by
\begin{align}
h_k &= \t A_k + \t B_k \int \f{d\eta}{z_T^2} \nn \\ &
= \t A_k + \t B_k \int \f{dt}{a(t)^3} \left(1 - \f{2 \rho}{\rho_c}\right),
\end{align}
and, when the equation of state is constant and the scale factor is given by \eqref{lqc-a}, the integral can be evaluated, giving
\begin{align} \label{lqc-sol-h}
h_k^q = \t C_k + \t D_k \Bigg[ & \f{\alpha t^3}{3} \,
{}_2F_1 \! \left( \tf{3}{2}, \tf{2+\om}{1+\om}; \tf{5}{2}; -\alpha t^2 \right) \nn \\ &
- t \, {}_2F_1 \! \left( \tf{1}{2}, \tf{2+\om}{1+\om}; \tf{3}{2}; -\alpha t^2 \right) \Bigg].
\end{align}

As in Sec.~\ref{s.sc-bounce}, the LQC solution can now be used in order to calculate how the solution from the contracting branch of the universe \eqref{class-sol-h} evolves through the bounce.  To do this, the first step is to choose $\t C_k$ and $\t D_k$ such that the LQC solution \eqref{lqc-sol-h} agrees with the classical solution in the regime where quantum gravity effects are negligible, namely when $t \ll -1 / 6 \pi G \rho_c$.  In this regime \cite{Abramowitz-Stegun}
\begin{align}
h_k^q \to & \,
\t C_k - \t D_k \left( \f{2 \om}{1+\om} \right) \f{\sqrt\pi \Gamma(\tf{2+\om}{1+\om}-\tf{3}{2})}{4 \sqrt\alpha \Gamma(\tf{2+\om}{1+\om})} \nn \\ &
- \f{\t D_k}{\alpha^{1/(1+\om)}} \left(\f{1+\om}{\om-1}\right) |t|^{(\om-1)/(1+\om)}.
\end{align}
and requiring that this expression match \eqref{class-sol-h} fully determines $\t C_k$ and $\t D_k$ in terms of $\t A_k$ and $\t B_k$.  Note once again that the terms written above are those that are relevant in order to match the LQC solution to the classical one, and that for $\om < -1/3$ there exist subleading contributions to the constant term which decay slower than $|t|^{(\om-1)/(1+\om)}$ as $t \to -\infty$.  However, these terms are not relevant in the matching between the classical solution \eqref{class-sol-h} and the LQC solution \eqref{lqc-sol-h}.

Requiring that the solutions \eqref{lqc-sol-h} and \eqref{class-sol-h} match for $t \ll -1 / 6 \pi G \rho_c$ gives the relations
\be
\t C_k = \t A_k - \t B_k \, \alpha^{\tf{1-\om}{2(1+\om)}} \cdot \f{2\om (\om-1)}{(1+\om)^2} \cdot \f{\sqrt\pi \, \Gamma(\tf{2+\om}{1+\om}-\tf{3}{2})}{4 \, \Gamma(\tf{2+\om}{1+\om})},
\ee
and
\be
\t D_k = \t B_k\,  \alpha^{1/(1+\om)} \left(\f{\om-1}{1+\om} \right).
\ee
From this, it is easy to verify that the post-bounce $t \gg 1 / 6 \pi G \rho_c$ form of $h_k$, keeping only the dominant constant term, is given by
\be
h_k^+ = \t C_k + \t D_k \left(\f{2 \om}{1+\om}\right) \f{\sqrt\pi \, \Gamma(\tf{2+\om}{1+\om}-\tf{3}{2})}{4 \sqrt\alpha \, \Gamma(\tf{2+\om}{1+\om})},
\ee
which in terms of $\t A_k$ and $\t B_k$ is
\be
h_k^+ = \t A_k - \t B_k \, \alpha^{\tf{1-\om}{2(1+\om)}} \cdot \f{2\om (\om-1)}{(1+\om)^2} \cdot \f{\sqrt\pi \, \Gamma(\tf{2+\om}{1+\om}-\tf{3}{2})}{2 \, \Gamma(\tf{2+\om}{1+\om})}.
\ee

This expression can further be simplified by considering the dynamics of the tensor perturbations in the contracting branch, exactly as was done for the scalar perturbations.  For an equation of state $-1 < \om < 1$, in the contracting branch the time-dependent term grows as $t \to 0$ and therefore quickly dominates over the constant term; thus in this case the constant term is negligible and $\t A_k$ can be set to zero.  On the other hand, for $\om > 1$ the time-dependent term in a contracting space-time is a decaying mode and it is the constant term that dominates as the bounce is approached: in this case it is the time-dependent mode that is negligible and $\t B_k$ that can be set to zero.

Therefore, the form of $h_k$ after the bounce, for $-1 < \om < 1$ is
\be
h_k^+ = - \t B_k \, \alpha^{\tf{1-\om}{2(1+\om)}} \cdot \f{2\om (\om-1)}{(1+\om)^2} \cdot \f{\sqrt\pi \, \Gamma(\tf{2+\om}{1+\om}-\tf{3}{2})}{2 \, \Gamma(\tf{2+\om}{1+\om})},
\ee
and for $\om > 1$
\be
h_k^+ = \t A_k.
\ee
Once again, the analysis can easily be extended for $\om=1$, with results exactly analogous to \eqref{om1-bef} and \eqref{om1-aft},
\be
h_k^- = A_k + B_k \ln |t| \quad \to \quad
h_k^+ = A_k - B_k \ln |t|.
\ee
Also, again as before, note that in the expanding branch there is a growing mode if $\om \ge 1$ that at sufficiently late times will dominate over the constant mode.  However, so long as the transition to the standard radiation-dominated cosmology occurs a relatively short time after the bounce, the dominant contribution to $h_k$ after the bounce ---even when $\om \ge 1$--- is the constant mode that has been considered here.

\subsection{The Tensor-to-Scalar Ratio}
\label{ss.r}

Recent observations of the CMB have determined strong bounds on the tensor-to-scalar ratio, which compares the amplitude of the power spectra of tensor and scalar perturbations,
\be
r = 64 \pi G \, \f{|h_k|^2}{|\mR_k|^2},
\ee
with the current upper bound on the tensor-to-scalar ratio being $r < 0.12~(95\%$ CL) \cite{Ade:2015tva} which rules out a large number of cosmological scenarios, including some inflationary models as well as some alternatives to inflation.  Interestingly, it has previously been shown in LQC that in some cases the tensor-to-scalar ratio can be significantly suppressed during the LQC bounce \cite{WilsonEwing:2012pu, Cai:2014jla}, thus providing some wiggle room to some cosmological models that would otherwise be ruled out \cite{Quintin:2015rta}.

With the results obtained in this paper, it is now possible to determine precisely which conditions are necessary for $r$ to be suppressed during the LQC.  First, it is important to recall that for $-1 < \om < 1$, $A_k$ and $\t A_k$ are negligible compared to $B_k$ and $\t B_k$ respectively, and that for $\om > 1$ it is the converse that is true.

Then, denoting the value of the tensor-to-scalar ratio at times well before the bounce by $r_-$ and its value at times well after the bounce by $r_+$, for $-1 < \om < 1$,
\be
r_- = 64 \pi G \, \f{|\t B_k|^2}{|B_k|^2},
\ee
and
\be
r_+ = 64 \pi G \left( \f{2 \om}{1+\om} \right)^2 \f{|\t B_k|^2}{|B_k|^2},
\ee
while for $\om > 1$,
\be
r_- = 64 \pi G \, \f{|\t A_k|^2}{|A_k|^2},
\ee
and
\be
r_+ = 64 \pi G \, \f{|\t A_k|^2}{|A_k|^2}.
\ee

Finally, combining these results (and including the $\om=1$ case) gives
\be
r_+ = \begin{cases}
\f{4 \om^2}{(1+\om)^2} \, r_- \quad &{\rm for~} -1 < \om \le 1, \\
r_- \quad &{\rm for~} \om \ge 1.
\end{cases}
\ee
In addition to recovering the previously known results that the tensor-to-scalar ratio is completely suppressed during the LQC bounce if $\om=0$ \cite{WilsonEwing:2012pu}, and that in the case of $\om=1/3, r_+ = r_- / 4$ \cite{Cai:2014jla}, this result shows how the LQC bounce affects the tensor-to-scalar ratio for any constant equation of state for the matter field.

In particular, for $\om \in [-1/3, 1]$, the tensor-to-scalar ratio is suppressed during the bounce, and by at least a factor of 1/4 if $\om \in [-1/5, 1/3]$.  The closer that the equation of state during the bounce is to zero, the more $r$ is suppressed by quantum gravity effects, and $r_+ \to 0$ as $\om \to 0$.

On the other hand, for $\om>1$, the tensor-to-scalar ratio is unaffected by the bounce, and for $\om \in (-1, -1/3)$, the tensor-to-scalar ratio is actually amplified during the bounce.  In fact, as $\om \to -1$, $r_+$ becomes arbitrarily large.

\section{Discussion}
\label{s.disc}

Using the separate universe approximation where long-wavelength perturbations in cosmology are modeled as a collection of super-horizon homogeneous patches that evolve independently, it is possible to obtain a full loop quantization of both the background FLRW space-time and long-wavelength scalar perturbations simply by performing a standard loop quantization in each homogeneous patch.  Then, for states where the LQC wave function in each patch is sharply peaked, the LQC effective Friedmann equations provide an excellent approximation to the full quantum dynamics in each patch.  From the LQC effective Friedmann equations, it is possible to extract the LQC effective equations of motion for the long-wavelength scalar perturbations via the separate universe framework, and then these effective equations can be used to calculate the evolution of long-wavelength scalar perturbations through the LQC bounce in the cases where (i) quantum fluctuations in the background FLRW space-time are negligible, (ii) the long-wavelength modes have been sufficiently squeezed so that the quantum fluctuations in the perturbations are also negligible, (iii) there are no entropy perturbations, and (iv) there is no anisotropic stress in the matter fields.  (If quantum fluctuations are important it is necessary to use the more complicated quantum equations of motion given in \cite{WilsonEwing:2012bx}.  In order to include entropy perturbations or matter fields with non-vanishing anisotropic stress, it will be necessary to extend these results or use a different framework.)

These results can be applied in two main settings.  First, there are some alternatives to inflation ---like the matter bounce scenario and the ekpyrotic universe--- where scale-invariant perturbations are generated in a contracting pre-bounce cosmology, and then the procedure described in this paper explains how the evolution of these perturbations across the LQC bounce can be calculated, which will determine the amplitude and scale-dependence of the perturbations after the bounce which can then be compared to observations.  Second, for the case of inflation, it is possible to calculate the correct form of the long-wavelength cosmological perturbations at the onset of inflation if the form of $\mR_k$ is known before the bounce.

The propagation of the curvature perturbations across the LQC bounce has already been calculated in a number of interesting contexts, including some realizations of the matter bounce scenario \cite{WilsonEwing:2012pu, Cai:2014zga, Cai:2014jla}, the ekpyrotic universe \cite{Wilson-Ewing:2013bla}, and in a pre-inflationary setting \cite{Linsefors:2012et}, and these results have now been extended in Sec.~\ref{s.sc-bounce} for all cases when the equation of state $\om$ is constant during the bounce.  Furthermore, it is clear how to generalize these results to allow for a dynamical $\om(t)$.  In addition, the evolution of the tensor modes across the bounce can also be calculated using the effective equations derived in \cite{Cailleteau:2012fy} using a different approach to cosmological perturbation theory based on an effective Hamiltonian.

A particularly interesting result is that the tensor-to-scalar ratio $r$ can be suppressed during the LQC bounce, in some cases significantly.  Indeed, for $\om \in [-1/5, 1/3]$, $r$ is suppressed by at least a factor of 1/4 during the bounce, and as $\om \to 0$, the value of the tensor-to-scalar ratio after the bounce goes to zero also.  On the other hand, if $\om < -1/3$, $r$ is amplified during the bounce.

This result is particularly important for alternatives to inflation where scale-invariant perturbations are generated in a pre-bounce contracting universe.  Then, by calculating the value of the tensor-to-scalar ratio before the bounce, it is possible to constrain the value of the equation of state during the bounce via the observational bound of $r < 0.12$.  If the matter fields dominating the dynamics at the bounce are known, then it may be possible to rule out cosmological scenarios based on the predicted value of $r$ after the bounce.

Finally, note that this suppression of $r$ during the bounce is an effect coming directly from LQC: if in the future observations favour a cosmological model where the scale-invariant perturbations are generated in a contracting pre-bounce phase but the tensor-to-scalar ratio is lower than expected from the pre-bounce dynamics, this lower value of $r$ could be due to quantum gravity effects and a signature of LQC.

\acknowledgments

This work was supported by a grant from the John Templeton Foundation.

%\newpage

%\bibliographystyle{bib-style}
%\bibliography{bibliography}

%\end{document}

\raggedright

\end{document}